\documentclass[a4paper,11pt]{article}

\usepackage[left=2.5cm,right=2.5cm,top=2.5cm,bottom=2.5cm]{geometry}
\usepackage{graphicx,amssymb,amsmath,amsthm,mathrsfs,setspace,subcaption,cite,authblk}
\usepackage[all]{xy}

\usepackage[colorlinks=true,bookmarks=true]{hyperref}
\usepackage{breakurl} 

\usepackage{mathtools}

\DeclarePairedDelimiterX\braket[2]{\langle}{\rangle}{#1 \delimsize\vert #2}

\hypersetup{citecolor=blue}
\newcommand{\dif}{\mathrm{d}}
\newcommand{\Eqref}[1]{(\ref{#1})}
\newcommand{\half}{\frac{1}{2}}

\newcommand{\brac}[1]{\left(#1 \right)}
\newcommand{\sbrac}[1]{\left[#1\right]}

\numberwithin{equation}{section}
 
\begin{document}

\title{Motion of charged particles in spacetimes with magnetic fields of spherical and hyperbolic symmetry}

\author{Yen-Kheng Lim\footnote{Email: yenkheng.lim@gmail.com, yenkheng.lim@xmu.edu.my}}

\affil{\normalsize{\textit{Department of Physics, Xiamen University Malaysia, 43900 Sepang, Malaysia}}}

\date{\normalsize{\today}}
\maketitle 
 
\renewcommand\Authands{ and }
\begin{abstract}
 The motion of charged particles in spacetimes containing a submanifold of constant positive or negative curvature is considered, with the electromagnetic tensor proportional to the volume two-form form of the submanifold. In the positive curvature case, this describes spherically symmetric spacetimes with a magnetic monopole, while in the negative curvature case, it is a hyperbolic spacetime with magnetic field uniform along hyperbolic surfaces. Constants of motion are found by considering Poisson brackets defined on a phase space with gauge-covariant momenta. In the spherically-symmetric case, we find a correspondence between the trajectories on the Poincar\'{e} cone with equatorial geodesics in a conical defect spacetime. In the hyperbolic case, the analogue of the Poincar\'{e} cone is defined as a surface in an auxiliary Minkowski spacetime. Explicit examples are solved for the Minkowski, $\mathrm{AdS}_4\times S^2$, and the hyperbolic AdS-Reissner--Nordstr\"{o}m spacetimes.
\end{abstract}

\section{Introduction} \label{intro}

An important aspect in the study of the motion of charged particles under Lorentz forces is the existence of constants of motion. If such quantities can be found, then the accessible regions of the particle's configuration space can be clearly identified, making the analysis of the problem tractable.

A central example for this paper is the \emph{Poincar\'{e} cone} \cite{Poincare1896} of a non-relativistic charged particle moving in the field of a magnetic monopole \cite{Schwinger:1976fr,Feher:1987vs,Sivardiere2000,Balian:2005joa,Mayrand:2014kja,Kupriyanov:2018xji,Marmo:2019ulx}. The cone arises from the $SO(3)$ symmetry of the system, leading to conserved quantities which can then be used to show that the trajectories of the particles are confined to a cone. A subclass of this problem is that of charged particles on a unit sphere with a constant magnetic field \cite{vanHolten:2006xq}. 

In another class of problems, one consideres the motion of charged particles confined on a hyperbolic plane, in the presence of magnetic field that is uniform over the plane. Such problems admit $SO(1,2)$ symmetry, and the equations of motion reveal two sets of solutions. For `strong' magnetic fields, the particles move along circles in the Poincar\'{e} half plane. While for `weak' magnetic field the circles intersect the boundary of the half plane. That is, the motion is unbounded and the particle escapes to infinity \cite{Comtet:1984mm,Comtet:1986ki,Kordyukov_2022}. 

In this paper, we wish to consider relativistic versions of the problems mentioned above. In particular, in the first class of problems we consider charged particles in spacetimes with $SO(3)$ isometry. That is, the spacetime metric contains a sphere $S^2$ as a submanifold. The magnetic field is given by a Maxwell tensor that is proportional to the volume form of $S^2$ and describes a field due a magnetic monopole. Despite the spacetime being curved, we can still obtain a Poincar\'{e} cone if an appropriate radial coordinate $r$ can be taken together with the angular coordinates on $S^2$ to describe the spherical coordinate system of an auxiliary Euclidean space $\mathbb{R}^3$. Then the trajectories lie on a Poincar\'{e} cone residing in this auxiliary space. 

Particle motion with Poincar\'{e} cones have previously been studied, for instance, around magnetically-charged Reissner--Nordstr\"{o}m spacetimes \cite{Grunau:2010gd,Pugliese:2011py,2017IJMPD..2650091S}, and black hole/solitons with a compact extra dimension stabilised by a magnetic flux \cite{Lim:2021ejg}. Recently, shadows and null geodesics around this latter system has been studied as well \cite{Guo:2022nto}. In this paper, we study the \emph{generic} situation of any spacetime with spherical symmetry with a magnetic monopole field. It will be shown that the full analytical solution in the spherical section applies to all spacetimes with the requisite properties. We will also establish a correspondence between charged particle motion on a Poincar\'{e} cone with \emph{geodesics}\footnote{To be precise, by \emph{geodesics} we refer to trajectories obeying the spacetime geodesic equation $\ddot{x}^\mu+\Gamma^\mu_{\alpha\beta}\dot{x}^\alpha\dot{x}^\beta=0$, as opposed to \emph{charged particle motion with Lorentz forces} which obeys $\ddot{x}^\mu+\Gamma^\mu_{\alpha\beta}\dot{x}^\alpha\dot{x}^\beta=e{F^\mu}_\nu\dot{x}^\nu$.} on spacetimes with a conical singularity.

In the second class of problems, we consider spacetimes with $SO(2,1)$ isometry. For this case the spacetime metric contains a hyperbolic submanifold $H^2$ of constant negative curvature. The magnetic field is a Maxwell tensor proportional to the volume form of $H^2$ and describes a  field that is constant along hyperbolic sections of the spacetime. The $SO(2,1)$ symmetries provide the constants of motion which then leads to an analogue of the Poincar\'{e} cone. However, $H^2$ cannot be embedded in $\mathbb{R}^3$. Rather, it has a natural embedding in Minkowski spacetime $\mathbb{R}^{2,1}$, therefore the $H^2$-analogue of the Poincar\'{e} cone lives in a spacetime with Lorentzian signature. 

The rest of paper is organised as follows. In Sec.~\ref{sec_symmetry} we outline a general procedure to find constants of motion. Magnetic fields in spherical symmetry are considered in Sec.~\ref{sec_spherical}, and those with hyperbolic symmetry are considered in Sec.~\ref{sec_hyperbolic}. Conclusions and closing remarks are given in Sec.~\ref{sec_conclusion}. We work in units where the speed of light is $c=1$ and our convention for Lorentzian signature is $(-,+,\ldots,+)$.

\section{Symmetries and Killing vectors} \label{sec_symmetry}

Let us begin in a general setting of an $(n+2)$-dimensional spacetime $M$, described in local coordinates with the metric $\dif s^2=g_{\mu\nu}\dif x^\mu\dif x^\nu$. The spacetime carries an electromagnetic field $F=\dif A$, where $A$ is the one-form potential $A=A_\mu\dif x^\mu$. A charged particle moving in this spacetime is described by the Lagrangian
\begin{align}
 \mathcal{L}=\half g_{\mu\nu}\dot{x}^\mu\dot{x}^\nu+eA_\mu\dot{x}^\mu,
\end{align}
where over-dots denote derivatives with respect to an affine parameter $\tau$, scaled such that, for time-like particles, we have $g_{\mu\nu}\dot{x}^\mu\dot{x}^\nu=-1$ along its motion. The parameter $e$ denotes the charge per unit mass of the particle. The canonical momenta is obtained form the Lagrangian by 
\begin{align}
 p_\mu=\frac{\partial\mathcal{L}}{\partial\dot{x}^\mu}=g_{\mu\nu}\dot{x}^\nu+eA_\mu.
\end{align}
We pass to the Hamiltonian by performing the Legendre transform $H=p_\mu\dot{x}^\mu-\mathcal{L}=\half g^{\mu\nu}\brac{p_\mu-eA_\mu}\brac{p_\nu-eA_\nu}$. However, in the presence of Lorentz interaction, it is often convenient to work with the gauge-covariant momenta $P_\mu=p_\mu-eA_\mu$ \cite{Feher:1987vs,NovikovBook,vanHolten:2006xq}. Then the Hamiltonian simply reads
\begin{align}
 H=\half g^{\mu\nu}P_\mu P_\nu. \label{Hamiltonian}
\end{align}
The phase space $\mathcal{P}$ is then described by the coordinates $(x^\mu,P_\nu)$. The symplectic form on $\mathcal{P}$ is \cite{NovikovBook}
\begin{align}
 \Omega=\dif x^\mu\wedge\dif P_\mu-\frac{e}{2}F_{\mu\nu}\dif x^\mu\wedge\dif x^\nu.
\end{align}
Given an observable $f:\mathcal{P}\rightarrow\mathbb{R}$, we define its associated Hamiltonian vector field $X_f$ by
\begin{align}
 \mathbf{i}_{X_f}\Omega=\dif f,
\end{align}
where $\mathbf{i}$ denotes the contraction $\mathbf{i}_X\Omega=\Omega(X,\cdot)$. The Poisson bracket of two observables $f$ and $g$ is defined by $\{f,g\}=\Omega(X_f,X_g)$. In coordinates, it is
\begin{align}
 \{f,g\}=\frac{\partial f}{\partial x^\mu}\frac{\partial g}{\partial P_\mu}-\frac{\partial f}{\partial P_\mu}\frac{\partial g}{\partial x^\mu}+eF_{\mu\nu}\frac{\partial f}{\partial P_\mu}\frac{\partial g}{\partial P_\nu}.
\end{align}
In particular, we have the following commutation relations for the phase-space coordinates
\begin{align*}
 \{x^\mu,x^\nu \}=0,\quad \{x^\mu,P_\nu\}=\delta^\mu_\nu,\quad \{P_\mu,P_\nu\}=eF_{\mu\nu}.
\end{align*}

In the following we use Poisson brackets to search for constants of motion for charged particles described by a Hamiltonian of the form \Eqref{Hamiltonian}. Then an observable $Q$ is a constant of motion if it Poisson-commutes with the Hamiltonian, $\{Q,H\}=0$. Using van Holten's prescription \cite{vanHolten:2006xq}, let us suppose that the constants of motion are linear in the momenta. That is, $Q$ is written as
\begin{align}
 Q=\Psi+\xi^\mu P_\mu,
\end{align}
where the scalar function $\Psi$ and vector $\xi^\mu$ depend only on the position variables $x^\mu$. Then, finding such a $Q$ amounts to finding $\Psi$ and $\xi^\mu$ such that $\{Q,H \}=0$. This leads to
\begin{align}
 g^{\mu\nu}P_\nu\brac{\partial_\mu\Psi+eF_{\lambda\mu}\xi^\lambda}+P_\mu P_\nu\nabla^{(\mu}\xi^{\nu)}=0. \label{Qeqn1}
\end{align}
If the $\xi^\mu$ are the components of a Killing vector $\xi=\xi^\mu\partial_\mu$, the second term of Eq.~\Eqref{Qeqn1} vanishes by Killing's equation. What remains is to solve
\begin{align}
 \partial_\mu\Psi+eF_{\lambda\mu}\xi^\lambda=0.
\end{align}
In the language of differential forms, the above equation can be written as $\dif\Psi+e\mathbf{i}_\xi F=0$.

In this paper we wish to consider spacetimes with magnetic fields that are uniform over geometries with constant curvature. Spherically symmetric magnetic fields (magnetic monopoles) are uniform on constant-radius spherical surfaces, and its Maxwell two-form $F$ is proportional to a volume form on $S^2$. Similarly, magnetic fields that are uniform over a flat plane or hyperbolic plane has $F$ proportional to the volume form of zero or negative curvature, respectively. 

To this end express our spacetimes in the form of warped products of $\mathcal{M}_n\times\mathcal{N}_2$. Taking $x^\alpha=(x^1,\ldots,x^n)$ to be coordinates on $\mathcal{M}_n$ and $y^a=(y^1,y^2)$ to be coordinates on $\mathcal{N}_2$, the metric is 
\begin{align}
 \dif s^2&=h_{\alpha\beta}\dif x^\alpha\dif x^\beta+\mathcal{C}(x)\bar{g}_{ab}\dif y^a\dif y^b, \label{metric}
\end{align}
where $\mathcal{C}(x)$ is a function of $x^\alpha=(x^1,\ldots,x^n)$ only. The electromagnetic fields are then written as $F=B\omega$, where $\omega$ is the volume two-form on $\mathcal{N}_2$ and $B$ is a constant parameter. In coordinates, 
\begin{align}
 F_{\mu\nu}=\left\{\begin{array}{cc}
                   B \sqrt{|\bar{g}|}\epsilon_{ab}, & \mbox{if } \mu\nu=ab,\\
                   0, & \mbox{otherwise},
                   \end{array}\right.
\end{align}
where $\bar{g}\equiv\det(\bar{g}_{ab})$ is the determinant of the metric on $\mathcal{N}_2$. This form solves the Maxwell equation on the spacetime.  

If $\mathcal{N}_2$ is a constant-curvature space, it is maximally symmetric and therefore has three independent Killing vector fields. Let us denote them by $\xi_{(i)}=\xi_{(i)}^a\partial_a$ for $i=1,2,3$. In this situation, Eq.~\Eqref{Qeqn1} reads 
\begin{align}
 \partial_a\Psi_{(i)}+eB\sqrt{|\bar{g}|}\epsilon_{ba}\xi^b_{(i)}=0,\quad i=1,2,3, \label{Qeqn2}
\end{align}
or, expressed as differential forms,
\begin{align}
 \dif\Psi_{(i)}+eB\mathbf{i}_{\xi_{(i)}}\omega=0.
\end{align}
The existence of a solution $\Psi$ to this equation is equivalent to the statement of whether the one-form $\mathbf{i}_{\xi_{(i)}}\omega$ is exact. Now, if $\mathcal{N}_2$ has a finite fundamental group, any one-form on it is exact.\footnote{The author thanks Mounir Nisse for pointing this out.} In fact, we are considering constant-curvature spaces for $\mathcal{N}_2$, which are simply connected. Therefore a solution $\Psi_{(i)}$ exists, and therefore we have three constants of motion
\begin{align}
 Q_{(i)}=\Psi_{(i)}+\xi_{(i)}^aP_a,
\end{align}
for $i=1,2,3$.

Suppose that the gauge potential is the form $A=A_a\dif y^a$ such that $F=\dif A=B\omega$, where the components $A_a$ depend only on coordinates $y^a$ only (and not on any $x^\alpha$). Then, in terms of the canonical momenta, the Hamilton--Jacobi equation under the metric \Eqref{metric} reads 
\begin{align}
 \half\sbrac{h^{\alpha\beta}\frac{\partial S}{\partial x^\alpha}\frac{\partial S}{\partial x^\beta}+\frac{1}{\mathcal{C}}\bar{g}^{ab}\brac{\frac{\partial S}{\partial y^a}-eA_a}\brac{\frac{\partial S}{\partial y^b}-eA_b}}+\frac{\partial S}{\partial\tau}=0. \label{general_HJE}
\end{align}
This can be partially separated by the ansatz $S=\half\tau+\Phi(x^1,\ldots,x^n)+W(y^1,y^2)$. This leads to a separation constant $K$ such that
\begin{align}
 h^{\alpha\beta}\frac{\partial\Phi}{\partial x^\alpha}\frac{\partial\Phi}{\partial x^\beta}&=-\frac{K}{\mathcal{C}}-1,\label{eom_M}\\
 \bar{g}^{ab}\brac{\frac{\partial W}{\partial y^a}-eA_a}\brac{\frac{\partial W}{\partial y^b}-eA_b}&=K.\label{eom_N}
\end{align}
Thus the Hamilton--Jacobi equation decomposes into two separate problems. The first \Eqref{eom_M} is the equation for a Hamilton's characteristic function $\Phi$ for particle motion on $\mathcal{M}_n$, with an effective potential $K/\mathcal{C}+1$, and second \Eqref{eom_N} is equivalent to Hamilton's characteristic function $W$ of a \emph{non-relativistic} motion of a charged particle of energy $\half K$ on $\mathcal{N}_2$. When $\mathcal{N}_2$ has constant curvature, Eq.~\Eqref{eom_N} can be solved exactly, independent of \Eqref{eom_M}.

\section{Magnetic fields with spherical symmetry} \label{sec_spherical}

If $\mathcal{N}_2$ is a two-sphere $S^2$, its metric can be written in usual spherical coordinates as $\dif\Omega^2_{(2)}=\dif\theta^2+\sin^2\theta\,\dif\phi^2$. Therefore the metric \Eqref{metric} takes the form
\begin{align}
 \dif s^2&=h_{\alpha\beta}\dif x^\alpha\dif x^\beta+\mathcal{C}(x)\brac{\dif\theta^2+\sin^2\theta\,\dif\phi}.\label{metric_sph}
\end{align}
The Maxwell two-form is 
\begin{align}
 F=B\sin\theta\,\dif\theta\wedge\dif\phi, \label{F_sph}
\end{align}
which comes from the exterior derivative of potential $A=-B\cos\theta\,\dif\phi$. The isometries on this space are generated by the Killing vectors 
\begin{subequations} \label{C_sph}
\begin{align}
 \xi_{(1)}&=-\sin\phi\partial_\theta-\cot\theta\cos\phi\partial_\phi,\\
 \xi_{(2)}&=\cos\phi\partial_\theta-\cot\theta\sin\phi\partial_\phi,\\
 \xi_{(3)}&=\partial_\phi.
\end{align}
\end{subequations}
For $F$ and $\xi_{(i)}$ given by \Eqref{F_sph} and \Eqref{C_sph}, respectively, Eq.~\Eqref{Qeqn2} is solved for $\Psi_{(i)}$, we obtain the constants of motion $Q_{(i)}=\Psi_{(i)}+\xi_{(i)}^aP_a$. Explicitly, they are \cite{vanHolten:2006xq}
\begin{subequations}\label{Q_sph}
 \begin{align}
  Q_{(1)}&=-eB\sin\theta\cos\phi-P_\theta\sin\phi-P_\phi\cot\theta\cos\phi,\\
  Q_{(2)}&=-eB\sin\theta\sin\phi+P_\theta\cos\phi-P_\phi\cot\theta\sin\phi,\\
  Q_{(3)}&=-eB\cos\theta+P_\phi.
 \end{align}
\end{subequations}

\subsection{Analysis of charged-particle motion on \texorpdfstring{$\mathcal{N}_2=S^2$}{N2=S2}} \label{subsec_BaseS2}

In this subsection consider the intrinsic problem of a (non-relativistic) particle moving on a unit sphere $S^2$. This problem was considered studied by van Holten in \cite{vanHolten:2006xq}. Here, we will provide a detailed discussion of the problem, along with its exact solution. We will also review how the Poincar\'{e} cone is defined. 

The metric of a sphere of radius $a$ is given by
\begin{align}
 \dif s^2=a^2\brac{\dif\theta^2+\sin^2\theta\dif\phi^2},
\end{align}
and the magnetic potential is $A=-B\cos\theta\,\dif\phi$. The magnetic field strength $F=B\sin\theta\,\dif\theta\wedge\dif\phi$ is constant on the sphere.

The Lagrangian for the non-relativistic problem is 
\begin{align}
 \mathcal{L}=\frac{a^2}{2}\brac{\dot{\theta}^2+\sin^2\theta\dot{\phi}^2}-eB\cos\theta\dot{\phi}.
\end{align}
The canonical momenta are 
\begin{align}
 p_\theta=a^2\dot{\theta},\quad p_\phi=a^2\sin^2\theta\dot{\phi}-eB\cos\theta,
\end{align}
whereas the gauge-covariant momenta are 
\begin{align}
 P_\theta=p_\theta=a^2\dot\theta,\quad P_\phi=p_\phi+eB\cos\theta=a^2\sin^2\theta\dot{\phi}.
\end{align}
Its corresponding Hamiltonian is $H=\frac{1}{2a^2}\sbrac{p_\theta^2+\frac{\brac{p_\phi+eB\cos\theta}^2}{\sin^2\theta}}$, for which the Hamilton--Jacobi equation is
\begin{align}
 \frac{1}{2a^2}\sbrac{\brac{\frac{\partial S}{\partial\theta}}^2+\frac{1}{\sin^2\theta}\brac{\frac{\partial S}{\partial\phi}+eB\cos\theta}^2}+\frac{\partial S}{\partial\tau}=0.
\end{align}
Taking the separation ansatz to be $S=-\mathcal{E}\tau+L\phi+S_\theta(\theta)$, where we recognise $\mathcal{E}>0$ as the (non-relativistic) total energy and $W=L\phi+S_\theta(\theta)$ as the Hamilton's characteristic function, and $L$ is the angular momentum associated with $\phi$-rotations. With $p_\theta=a^2\dot{\theta}=\frac{\dif S_\theta}{\dif\theta}$, the equations of motion are 
\begin{subequations}
\begin{align}
 a^2\dot{\theta}&=\pm\sqrt{K-\frac{\brac{L+eB\cos\theta}^2}{\sin^2\theta}},\label{baseS2_thetadot}\\
 a^2\dot{\phi}&=\frac{L+eB\cos\theta}{\sin^2\theta},\label{baseS2_phidot}
\end{align}
\end{subequations}
where we have denoted $K=a^2\mathcal{E}$. This equation can be slightly simplified by letting $x=\cos\theta$, which leads to
\begin{subequations}
\begin{align}
 \dot{x}&=\mp\sqrt{X(x)},\label{S2_xdot}\\
 \dot{\phi}&=\frac{L+eBx}{1-x^2},\label{S2_phidot}\\
 X(x)&=-(K+e^2B^2)x^2-2eBLx+K-L^2.\label{S2_X}
\end{align}
\end{subequations}
From Eq.~\Eqref{S2_xdot}, the domain of allowed motion are the points $x$ where $X(x)\geq0$. Since $X(x)$ is a quadratic function with a negative quadratic coefficient, the condition $X(x)\geq0$ corresponds to a finite domain $x_-\leq x\leq x_+$, where 
\begin{align}
 x_\mp=\frac{-eBL\mp\sqrt{K\brac{K-L^2+e^2B^2}}}{K+e^2B^2}.
\end{align}
The notion of the Poincar\'{e} cone can be obtained as follows. We embed $S^2$ as a sphere of radius $r$ in $\mathbb{R}^3$ described by coordinates
\begin{align}
 X^1&=r\sin\theta\cos\phi,\quad X^2=r\sin\theta\sin\phi,\quad X^3=r\cos\theta.
\end{align}
The position vector of the particle is given by
\begin{align}
 \vec{r}=X^1\hat{e}_1+X^2\hat{e}_2+X^3\hat{e}_3,
\end{align}
where $\hat{e}_1$, $\hat{e}_2$, and $\hat{e}_3$ are the standard Cartesian basis of $\mathbb{R}^3$. 

\begin{figure}
 \centering
 \includegraphics[scale=0.4]{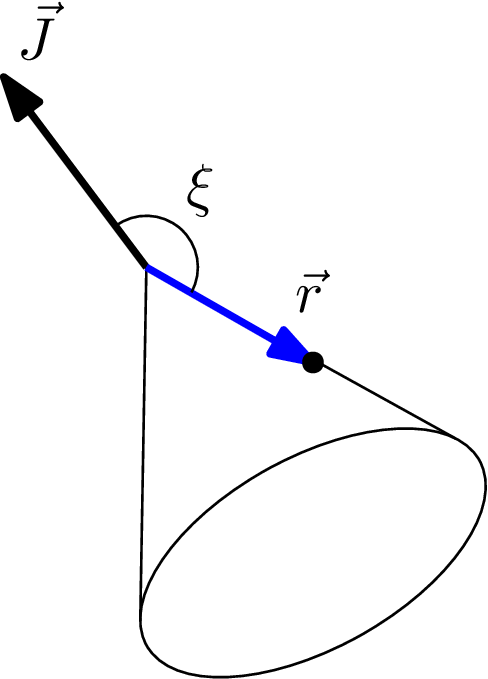}
 \caption{A sketch of the Poincar\'{e} cone, where $\vec{J}$ is given by Eq.~\Eqref{sph_J}. The apex of the cone is the origin of $\mathbb{R}^3$, and the particle's trajectory on $S^2$ is a circle, shown here as the circle at the base of the cone.}
 \label{fig_PoincareCone}
\end{figure}

Let the constants of motion $Q_{(i)}$ given in Eq.~\Eqref{Q_sph} be the components of a vector 
\begin{align}
 \vec{J}=Q_{(1)}\,\hat{e}_2+Q_{(2)}\,\hat{e}_1+Q_{(3)}\,\hat{e}_3,\label{sph_J}
\end{align}
whose norm is $J=\sqrt{Q_{(1)}^2+Q_{(2)}^2+Q_{(3)}^2}=\sqrt{e^2B^2+K}$. Now, since each $Q_{(i)}$ are constants of motion, the vector $\vec{Q}$ is a fixed vector. Its standard dot product in $\mathbb{R}^3$ with the position vector leads to
\begin{align}
 \frac{\vec{J}}{J}\cdot\frac{\vec{r}}{r}=-\frac{eB}{\sqrt{e^2B^2+K}}.\label{sph_InnerProd}
\end{align}
In other words, the particle's position vector $\vec{r}$ is always at fixed angle $\xi=\arccos\brac{-\frac{eB}{\sqrt{e^2B^2+K}}}$ with $\vec{J}$. Now, the set of points whose position vector is at constant angle with a fixed vector $\vec{J}$ defines the Poincar\'{e} cone. However, in the present situation, the particle is moving on the sphere of constant $r=a$, the particle trajectory is the intersection of a cone and a sphere, which is a circle. The vector $\vec{J}$ is along a line perpendicular to the plane of the circle and passes from the circle's centre to the origin of $\mathbb{R}^3$, as shown in Fig.~\ref{fig_PoincareCone}.

We can make use of the spherical symmetry to align the axis of the coordinate system with $\vec{J}$. First, observe that the $SO(3)$ symmetry preserves the inner product \Eqref{sph_InnerProd}. This symmetry can be use to rotate the $X^3$-axis to align with $\vec{J}$. This amounts to having
\begin{align*}
 Q_{(1)}&=0=-\cos\phi\brac{eB\sin\theta+P_\phi\cot\theta}-P_\theta\sin\phi,\\
 Q_{(2)}&=0=-\sin\phi\brac{eB\sin\theta+P_\phi\cot\theta}+P_\theta\cos\phi.
\end{align*}
By Eq.~\Eqref{S2_xdot}, for $K=L^2-e^2B^2$, the coordinate $\cos\theta=x$ is constant at $x=-\frac{eB}{L}$, leading to $P_\theta=0$. For these parameters, the equation of motion for $\phi$ is simply $\dot{\phi}=L$. This subsequently leads to $eB\sin\theta+P_\phi\cot\theta=0$, giving $Q_{(1)}=Q_{(2)}=0$ as desired. This aligns $\vec{J}$ to be along the $X^3$-axis, and the cone subtends the half-angle $\xi=\arccos\brac{-\frac{eB}{|L|}}$.  

Next we give the full solutions to the equations of motion for arbitrary directions of $\vec{J}$. (Where it is not necesssarily oriented along $X^3$.) The following results has been solved in earlier works, such as \cite{Grunau:2010gd,Lim:2021ejg}, but in their respective contexts. Here, we shall give a detailed discussion to complete our study of generic charged particle motion in $S^2$ symmetry. 

\begin{figure}
 \centering
 \begin{subfigure}[b]{0.32\textwidth}
    \centering
    \includegraphics[width=\textwidth]{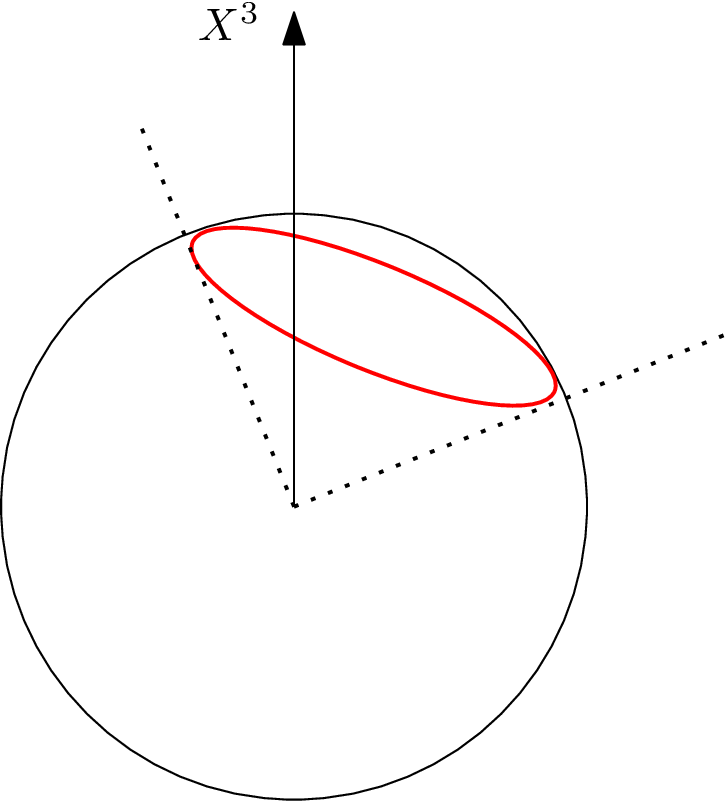}
    \caption{$\Delta\phi=2\pi$.}
    \label{fig_Cone_a}
  \end{subfigure}
  \begin{subfigure}[b]{0.32\textwidth}
    \centering
    \includegraphics[width=0.88\textwidth]{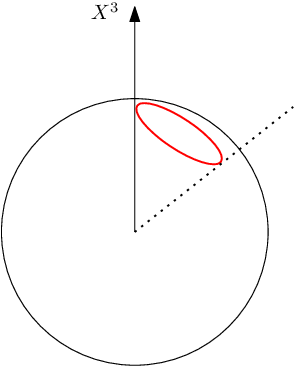}
    \caption{$\Delta\phi=\pi$.}
    \label{fig_Cone_b}
  \end{subfigure}
  \begin{subfigure}[b]{0.32\textwidth}
    \centering
    \includegraphics[width=0.88\textwidth]{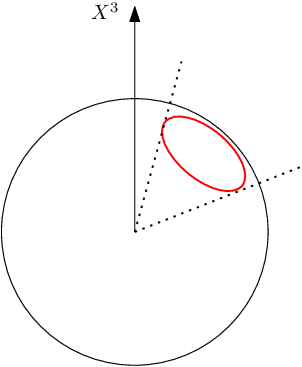}
    \caption{$\Delta\phi=0$.}
    \label{fig_Cone_c}
  \end{subfigure}
  \caption{Trajectory of charged particles on $S^2$. The periodicity $\Delta\phi$ depends on whether the cone encloses the $X^3$-axis. The case \ref{fig_Cone_a} occurs when $\mathrm{sgn}(L-eB)=\mathrm{sgn}(L+eB)$, where the $\phi$-evolution is a rotation. The case \ref{fig_Cone_c} is $\mathrm{sgn}(L-eB)=-\mathrm{sgn}(L+eB)$, where the $\phi$-evolution is a libration. The case \ref{fig_Cone_b} is the critical case, where the circle intersects the axis. The outline of the Poincar\'{e} cone is shown as the dotted lines.}
\end{figure}

Equations.~\Eqref{S2_xdot} and \Eqref{S2_phidot} leads to
\begin{align}
 \frac{\dif\phi}{\dif x}=\mp\frac{L+eBx}{(1-x^2)\sqrt{X(x)}}.
\end{align}
Suppose the particle starts at initial conditions $\phi=0$, $\dot{\phi}>0$, and $x=x_-$. Upon performing a partial fraction decomposition, for $x\leq x \leq x_+$,
\begin{align*}
 \phi(x)&=-\frac{L-eB}{2}\int_{x_+}^x\frac{\dif x'}{(1+x')\sqrt{X(x')}}-\frac{L+eB}{2}\int_{x_+}^x\frac{\dif x'}{(1-x)\sqrt{X(x')}}.
\end{align*}
The integrals can be evaluated exactly, giving
\begin{subequations} \label{baseS2_soln}
\begin{align}
 \phi(x)&=\half\mathrm{sgn}(L-eB)\brac{\zeta(x)+\frac{\pi}{2}}-\half\mathrm{sgn}(L+eB)\brac{\eta(x)-\frac{\pi}{2}},\quad x_-\leq x < x_+,\\
 \zeta(x)&=\arcsin\sbrac{\frac{1}{\sqrt{K(K-L^2+e^2B^2)}}\brac{\frac{(L-eB)^2}{1+x}-\brac{K+e^2B^2-eBL}}},\\
 \eta(x)&=\arcsin\sbrac{\frac{1}{\sqrt{K(K-L^2+e^2B^2)}}\brac{\frac{(L+eB)^2}{1-x}-\brac{K+e^2B^2+eBL}}}.
\end{align}
\end{subequations}
In particular, note that $\zeta(x_\pm)=\mp\frac{\pi}{2}$ and $\eta(x_\pm)=\pm\frac{\pi}{2}$. The sign function `$\mathrm{sgn}$' is defined as $\mathrm{sgn}(x)=+1$ if $x>0$, $\mathrm{sgn}(x)=-1$ if $x<0$, and $\mathrm{sgn}(0)=0$. The appearance of the sign functions here are due to the topology of the closed orbits in relation to the $X^3$-axis. To see this, consider the accumulated angle $\Delta\phi$ as the particle completes one period of $\theta$-evolution. This is computed as 
\begin{align}
 \Delta\phi&=2\phi(x_-)=\sbrac{\mathrm{sgn}(L-eB)+\mathrm{sgn}(L+eB)}\pi.
\end{align}
So, $\Delta\phi$ is either $0$, $\pi$, or $2\pi$, depending on the relative signs of $L-eB$ and $L+eB$. In particular, if $\mathrm{sgn}(L-eB)=\mathrm{sgn}(L+eB)$, we have $|\Delta\phi|=2\pi$. This means the trajectory encloses $X^3$ axis as shown in Fig.~\ref{fig_Cone_a}. In this case motion would be called a \emph{rotation} (with respect to the $X^3$-axis).  On the other hand, if $(L-eB)$ and $(L+eB)$ have opposite signs, $\Delta\phi=0$. In this case the motion is a \emph{libration} and the trajectory does not enclose the $X^3$-axis, as shown in Fig.~\ref{fig_Cone_c}. The intermediate case is either $(L-eB)=0$ or $(L+eB)=0$, for which $|\Delta\phi|=\pi$. This is where the trajectory intersects the $X^3$ axis, as shown in Fig.~\ref{fig_Cone_b}.

\subsection{Correspondence between the Poincar\'{e} cone and geodesics on conical defect spacetimes} \label{subsec_correspondenceS2}

The primary goal of this subsection is to establish a correspondence between trajectories of the following two problems. The first of which is charged particle motion under the metric and Maxwell tensor 
\begin{subequations} \label{4sph_metric}
\begin{align}
 \dif s^2&=-\mathcal{A}(r)\dif t^2+\mathcal{B}(r)\dif r^2+\mathcal{C}(r)\brac{\dif\theta^2+\sin^2\theta\,\dif\phi^2},\\
   F&=B\sin\theta\,\dif\theta\wedge\dif\phi,
\end{align}
\end{subequations}
where $\mathcal{A}$, $\mathcal{B}$, and $\mathcal{C}$ are functions of $r$. This is the four-dimensional case of Eq.~\Eqref{metric_sph} obtained by taking $h_{\alpha\beta}\dif x^\alpha\dif x^\beta=-\mathcal{A}\dif t^2+\mathcal{B}\dif r^2$. This form contains the Schwarzschild, Reissner--Nordstr\"{o}m, or generally charged dilaton black hole solutions for appropriate choices of $\mathcal{A}$, $\mathcal{B}$, and $\mathcal{C}$. It is well known that trajectories for such a problem lie on the \emph{Poincar\'{e} cone}.

The second problem is \emph{geodesic} motion in the absence of Lorentz interaction, in a spacetime with a conical defect described by 
\begin{align}
 \dif s^2&=-\mathcal{A}(r)\dif t^2+\mathcal{B}(r)\dif r^2+\mathcal{C}(r)\brac{\dif\theta^2+\eta^2\sin^2\theta\,\dif\phi^2}. \label{4def_metric}
\end{align}
We take the domain of $\phi$ as $\phi\in[0,2\pi)$. Then the spacetime carries a conical singularity for $\eta\neq1$. This spacetime can be interpreted as taking a regular spacetime, removing a wedge of angle $\delta=2\pi(1-\eta)$ and subsequently gluing together the resulting edges. Physically, this signals the presence of a cosmic string of mass per unit length $\mu$ at the location of the conical singularity, where $\delta=8\pi G_{\mathrm{N}}\mu$, with $G_{\mathrm{N}}$ being the gravitational constant. The geometry is then that of a cone. In this sense, this correspondence is perhaps intuitively unsurprising, as both problems involve some concept of cones. In any case, we make this relation precise by showing that the equations of motion of the two problems are the same.

To start, we consider the first problem. For charged particles moving in a gravitational and electromagnetic field described by \Eqref{4sph_metric}, the canonical momenta are 
\begin{subequations}
\begin{align}
 -p_t&=\mathcal{A}\dot{t},\\
  p_r&=\mathcal{B}\dot{r},\\
  p_\theta&=\mathcal{C}\dot\theta,\\
  p_\phi&=\mathcal{C}\sin^2\theta\,\dot{\phi}-eB\cos\theta.
\end{align}
\end{subequations}
To relate the constants of motion $Q_{(i)}$ to the equations of motion, we turn to the Hamilton--Jacobi equation 
\begin{align}
 \half\sbrac{-\frac{1}{\mathcal{A}}\brac{\frac{\partial S}{\partial t}}^2+\frac{1}{\mathcal{B}}\brac{\frac{\partial S}{\partial r}}^2+\frac{1}{\mathcal{C}}\brac{\frac{\partial S}{\partial\theta}}^2+\frac{1}{\mathcal{C}\sin^2\theta}\brac{\frac{\partial S}{\partial\phi}-eB\cos\theta}^2}+\frac{\partial S}{\partial\tau}=0.
\end{align}
We separate this equation with the ansatz $S=\half\tau-Et+L\phi+S_r(r)+S_\theta(\theta)$. This implies the existence of a separation constant $K$ which ultimately leads to the following set of first-order equations
\begin{subequations}\label{eom_4sph}
\begin{align}
 \dot{t}&=\frac{E}{\mathcal{A}},\\
 \mathcal{B}\dot{r}^2&=\frac{E^2}{\mathcal{A}}-\frac{K}{\mathcal{C}}-1,\\
 \dot{\phi}&=\frac{L+eB\cos\theta}{\mathcal{C}\sin^2\theta},\\
 \mathcal{C}^2\dot{\theta}^2&=K-\frac{(L+eB\cos\theta)^2}{\sin^2\theta}. \label{4sph_thetadot}
\end{align}
\end{subequations}
The relation of these quantities to the gauge-covariant momenta are 
\begin{align}
 P_t=-\mathcal{A}\dot{t},\quad P_r=\mathcal{B}\dot{r},\quad P_\theta=\mathcal{C}\dot{\theta},\quad P_\phi=\mathcal{C}\sin^2\theta\dot{\phi}.
\end{align}
With these formulas, one can check that if Eq.~\Eqref{eom_4sph} is obeyed, then each $Q_{(i)}$ are indeed constant along the trajectory. This can be verified by explicitly computing $\frac{\dif}{\dif\tau}Q_{(i)}$ and checking that it equals zero. 

To visualise the trajectories, we let $(r,\theta,\phi)$ define points on (an auxiliary) three-dimensional Euclidean space $\mathbb{R}^3$ by
\begin{align}
 X^1=r\sin\theta\cos\phi,\quad X^2=r\sin\theta\sin\phi,\quad X^3=r\cos\theta,
\end{align}
Similar to what we did in Sec.~\ref{subsec_BaseS2}, we let the $Q_{(i)}$'s in \Eqref{Q_sph} be the components of a vector
\begin{align}
 \vec{J}=Q_{(1)}\,\hat{e}_1+Q_{(2)}\,\hat{e}_2+Q_{(3)}\,\hat{e}_3. \label{conevec_J}
\end{align}

As discussed in Sec.~\ref{subsec_BaseS2}, we choose the coordinate system such that the $X^3$-axis aligns with the cone vector $\vec{J}$, so that $\theta$ is constant throughout the motion and 
\begin{align}
 K=L^2-e^2B^2,\quad \cos\theta=-\frac{eB}{|L|}.
\end{align}
For these values, the equations of motion are now
\begin{subequations} \label{4sph_eom2}
 \begin{align}
  \dot{t}&=\frac{E}{\mathcal{A}},\\
  \frac{\sqrt{L^2-e^2B^2}}{L}\dot{\phi}&=\frac{\sqrt{L^2-e^2B^2}}{\mathcal{C}},\\
  \dot{r}^2&=\frac{E^2}{\mathcal{A}\mathcal{B}}-\brac{\frac{L^2-e^2B^2}{\mathcal{C}}+1}\frac{1}{\mathcal{B}},\\
  \theta&=\xi=\arccos\brac{-\frac{eB}{|L|}}.
 \end{align}
\end{subequations}

Next, we now show a correspondence between trajectories along the Poincar\'{e} cone with equatorial geodesics of a conical-defect spacetime, where the metric is Eq.~\Eqref{4def_metric}. The present task is to consider \emph{geodesics} in this spacetime, which is either the trajectory of neutral particles or simply charged particles in the absence of an electromagnetic field. The corresponding Lagrangian is 
\begin{align*}
 \mathcal{L}=\half\brac{-\mathcal{A}\dot{t}^2+\mathcal{B}\dot{r}^2+\mathcal{C}\dot{\theta}^2+\mathcal{C}\eta^2\sin^2\theta\dot{\phi}^2},
\end{align*}
from which we obtain the canonical momenta
\begin{subequations}
\begin{align}
 -p_t&=\mathcal{A}\dot{t},\\
  p_r&=\mathcal{B}\dot{r},\\
  p_\theta&=\mathcal{C}\dot{\theta},\\
  p_\phi&=\mathcal{C}\eta^2\sin^2\theta\,\dot{\phi}.
\end{align}
\end{subequations}
The Hamilton--Jacobi equation reads
\begin{align}
 \half\sbrac{-\frac{1}{\mathcal{A}}\brac{\frac{\partial S}{\partial t}}^2+\frac{1}{\mathcal{B}}\brac{\frac{\partial S}{\partial r}}^2+\frac{1}{\mathcal{C}}\brac{\frac{\partial S}{\partial\theta}}^2+\frac{1}{\mathcal{C}\eta^2\sin^2\theta}\brac{\frac{\partial S}{\partial\phi}}^2}+\frac{\partial S}{\partial\tau}=0.  
\end{align}
This equation can be separated by the ansatz $S=\half\tau-\mathcal{E}t+\Phi\phi+S_r(r)+S_\theta(\theta)$, where $\mathcal{E}$ and $\Phi$ are constants. Substitution of this ansatz into the Hamilton--Jacobi equation leads to a separation constant $\mathcal{K}$ which then results in the following equations of motion
\begin{subequations}
 \begin{align}
  \dot{t}&=\frac{\mathcal{E}}{\mathcal{A}},\\
  \dot{\phi}&=\frac{\Phi}{\mathcal{C}\eta^2\sin^2\theta},\\
  \mathcal{B}\dot{r}^2&=\frac{\mathcal{E}^2}{\mathcal{A}}-1-\frac{\mathcal{K}}{\mathcal{C}},\\
  \mathcal{C}^2\dot{\theta}^2&=\mathcal{K}-\frac{\Phi^2}{\eta^2\sin^2\theta}. \label{4def_thetadot}
 \end{align}
\end{subequations}
We consider the case $\mathcal{K}=\frac{\Phi^2}{\eta^2}$. From Eq.~\Eqref{4def_thetadot}, we find that the motion lies on the plane $\theta=\frac{\pi}{2}=\mathrm{constant}$. In this case, the equations of motion reduces to
\begin{subequations}\label{4def_eom}
\begin{align}
 \dot{t}&=\frac{\mathcal{E}}{\mathcal{A}},\\
 \eta\dot{\phi}&=\frac{\Phi}{\mathcal{C}\eta},\\
 \dot{r}^2&=\frac{\mathcal{E}^2}{\mathcal{AB}}-\brac{\frac{\Phi^2}{\eta^2\mathcal{C}}+1}\frac{1}{\mathcal{B}},\\
 \theta&=\frac{\pi}{2}.
\end{align}
\end{subequations}
By comparing Eq.~\Eqref{4def_eom} with \Eqref{4sph_eom2}, we find that, aside from $\theta$ being at different constants, the equations for $t$, $r$, and $\phi$ are the same upon the identification 
\begin{align}
 L^2=\frac{\Phi^2}{\eta^4},\quad e^2B^2=\frac{\Phi^2(1-\eta^2)}{\eta^4},\quad E=\mathcal{E}. \label{map}
\end{align}
Hence, by appropriately aligning the coordinate system with the Poincar\'{e} cone, the trajectories on the cone can be mapped to equatorial geodesics around a spacetime threaded by a cosmic string.

\begin{figure}
 \centering 
 \begin{subfigure}[b]{0.49\textwidth}
    \centering
    \includegraphics[width=\textwidth]{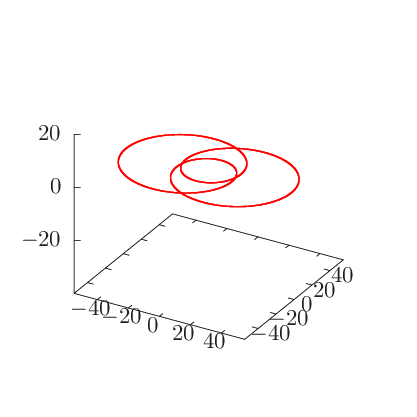}
    \caption{Neutral particle in Schwarzschild spacetime pierced with a cosmic string.}
    \label{fig_PeriodicPS}
  \end{subfigure}
  \begin{subfigure}[b]{0.49\textwidth}
    \centering
    \includegraphics[width=\textwidth]{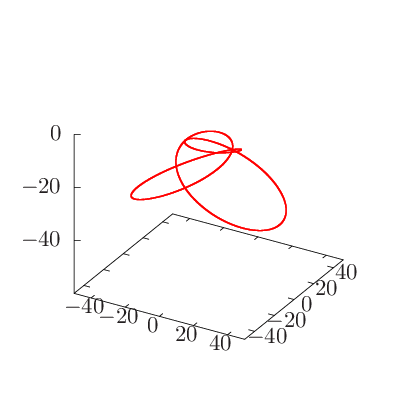}
    \caption{Charged particle in Schwarzschild spacetime with magnetic monopole.}
    \label{fig_PeriodicMS}
  \end{subfigure}
  \caption{In Fig.~\ref{fig_PeriodicPS}, the periodic orbit $(2,0,1)$ for the pierced Schwarzschild spacetime of $\eta=0.78$ is obtained for $\mathcal{E}=0.9855756508$, $\Phi=4$. Under the relation \Eqref{map}, we obtain the same orbit with $eB\simeq4.1142$, $L\simeq6.5746$, but it lies on the surface of a Poincar\'{e} cone. For these parameters, the cone surface is at angle $\xi=\arccos\brac{-\frac{eB}{|L|}}\simeq2.2469\;\mathrm{rad}$ from the positive $z$-axis, or $\simeq38.739^\circ$ under the $z=0$ plane.}
  \label{fig_Periodic}
\end{figure}

As an example, let us consider the Schwarzschild spacetime with a test magnetic monopole $F=B\sin\theta\,\dif\theta\wedge\dif\phi$, whose strength is sufficiently weak such that the field does not backreact to the spacetime curvature. As such the metric is given by $\mathcal{A}(r)=\frac{1}{\mathcal{B}(r)}=1-\frac{2m}{r}$ and $\mathcal{C}(r)=r^2$ giving the Schwarzschild metric with a monopole test magnetic field:
\begin{subequations}
\begin{align}
 \dif s^2&=-\brac{1-\frac{2m}{r}}\dif t^2+\brac{1-\frac{2m}{r}}^{-1}\dif r^2+r^2\brac{\dif\theta^2+\sin^2\theta\,\dif\phi^2},\\
   A&=-B\cos\theta\,\dif\phi.
\end{align}
\end{subequations}
The correspondence is with the Schwarzschild black hole threaded by a cosmic string \cite{Aryal:1986sz,Aliev:1988wv}, whose metric is
\begin{align}
 \dif s^2&=-\brac{1-\frac{2m}{r}}\dif t^2+\brac{1-\frac{2m}{r}}^{-1}\dif r^2+r^2\brac{\dif\theta^2+\eta^2\sin^2\theta\,\dif\phi^2}.
\end{align}
By the steps outlined above, the equations of motion for the two problems are identical under \Eqref{map}. Geodesics of the Schwarzschild spacetime with a cosmic string has previously been studied in Refs.~\cite{Galtsov:1989ct,Chakraborty:1991mb,Hackmann:2009rp}. As an explicit demonstration, Fig.~\ref{fig_PeriodicPS} shows a periodic orbit around a Schwarzschild spacetime threaded by a cosmic string with $\eta=0.78$. Specifically, this orbit labelled $(2,0,1)$ under Levin and Perez-Giz's taxonomy \cite{Levin:2008mq} and has energy $\mathcal{E}=0.9855756508$ and angular momentum $\Phi=4$. Under the map \Eqref{map} gives\footnote{We use the symbol `$\simeq$' to indicate that numerical values are given up to five significant figures.} $eB\simeq4.1142$ and $L\simeq6.5746$ we obtain a periodic orbit followed by a charged particle under the Lorentz force of a test magnetic monopole in a Schwarzschild background, shown in Fig.~\ref{fig_PeriodicMS}. It has the same topology $(2,0,1)$, but it lies on a cone which subtends an angle $2\xi=2\arccos\brac{-\frac{eB}{|L|}}$. 

\subsection{Charged particles in \texorpdfstring{$\mathrm{AdS}_n\times S^2$}{AdSn x S2} flux compactifications}

An example of a spacetime with $S^2$ symmetric magnetic field is the $\mathrm{AdS}_n\times S^2$ flux compactification. This is a particular case of a more general $\mathrm{AdS}_n\times S^q$ where the $q$-dimensional compactification is achieved by a $q$-form flux \cite{DeWolfe:2001nz}. Here, we consider the case $q=2$ which is described by the action
\begin{align}
 I=\frac{1}{16\pi G}\int\dif^{n+2}x\sqrt{-g}\brac{R-F^2}.
\end{align}
The Einstein--Maxwell equations are 
\begin{align}
 R_{\mu\nu}&=2F_{\mu\lambda}{F_\nu}^\lambda-\frac{1}{n}F^2g_{\mu\nu},\quad \nabla_\lambda F^{\lambda\mu}=0.
\end{align}
A solution describing $\mathrm{AdS}_n\times S^2$ is given by 
\begin{subequations}
\begin{align}
 \dif s^2&=-f(r)\dif t^2+f(r)^{-1}\dif r^2+r^2\dif\Omega^2_{(n-2)}+a^2\brac{\dif\theta^2+\sin^2\theta\,\dif\phi^2},\\
  A&=-B\cos\theta\,\dif\phi, \\
  f(r)&=1-\frac{\mu}{r^{n-3}}+\frac{nr^2}{2(n-1)^3B^2},\quad a=\sqrt{\frac{2(n-1)}{n}}|B|,
\end{align}
\end{subequations}
where $\dif\Omega^2_{(n-2)}$ is the metric of a unit $(n-2)$-sphere. In the case $n=2$, we take $\dif\Omega^2_{(n-2)}=0$, and the solution is the Bertotti--Robinson spacetime.

Particle motion in Bertotti--Robinson spacetime have been studied in \cite{Sokolowski:2014hia,Alekseev:2015hkr}. Various aspects of the $\mathrm{AdS}_{n}\times S^q$ and related compactifications have been studied in \cite{Freund:1980xh,DeWolfe:2001nz,Brown:2014sba}. As the external space is an Anti-de Sitter spacetime, particle motion might be of interest in the context of the AdS/CFT correspondence \cite{Berenstein:2020vlp}.
In the following let us consider black holes in the case $\mathrm{AdS}_4\times S^2$. In this case, 
\begin{subequations}
\begin{align}
 \dif s^2&=-f(r)\dif t^2+f(r)^{-1}\dif r^2+r^2\brac{\dif\zeta^2+\sin^2\zeta\,\dif\psi^2}+a^2\brac{\dif\theta^2+\sin^2\theta\,\dif\phi^2},\\
  f(r)&=1-\frac{2m}{r}+\frac{r^2}{4B^2},\quad a=\frac{\sqrt{6}}{2}|B|,
\end{align}
\end{subequations}
with the gauge potential still being $A=-B\cos\theta\,\dif\phi$. The Hamilton--Jacobi equation is \Eqref{general_HJE}, \Eqref{eom_M}, and \Eqref{eom_N} with
\begin{align*}
 \gamma_{\alpha\beta}\dif x^\alpha\dif x^\beta&=-f(r)\dif t^2+f(r)^{-1}\dif r^2+r^2\dif\zeta^2+\sin^2\zeta\,\dif\psi^2,\\
 \bar{g}_{ab}\dif\theta^a\dif\theta^b&=a^2\brac{\dif\theta^2+\sin^2\theta\,\dif\phi^2},
\end{align*}
Without loss of generality, we consider the motion confined in the $\zeta=\frac{\pi}{2}$ plane, for which the equations of motion are
\begin{subequations} \label{fluxcomp_eom}
 \begin{align}
  a^2\dot{\theta}&=\pm\sqrt{K-\frac{(L+eB\cos\theta)^2}{\sin^2\theta}}, \label{fluxcomp_eom_thetadot}\\
  a^2\dot{\phi}&=\frac{L+eB\cos\theta}{\sin^2\theta},\\
  \dot{t}&=-\frac{E}{f}, \label{fluxcomp_eom_phidot}\\
  \dot{\psi}&=\frac{\Psi}{r^2},\\
  \dot{r}&=\pm\sqrt{E^2-\brac{\frac{4K}{6B^2}+\frac{\Psi^2}{r^2}+1}f(r)}, \label{fluxcomp_eom_rdot}
 \end{align}
\end{subequations}
where $E$ and $\Psi$ are the conserved energy and angular momentum associated to the symmetry in the $t$ and $\psi$ directions, respectively.

\begin{figure}
 \centering 
 \includegraphics{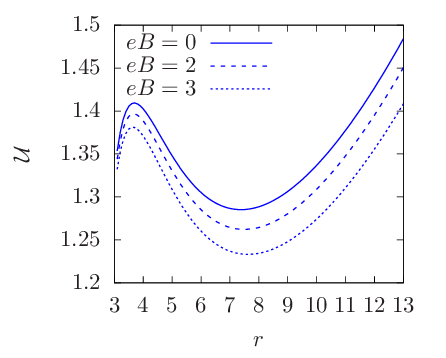}
 \caption{$m=1$, $B=10$, $L=2$, $\Psi=5$, $K=L^2-q^2$, and various $q$.}
 \label{fig_fluxcomp_pot}
\end{figure}

Equations \Eqref{fluxcomp_eom_thetadot} and \Eqref{fluxcomp_eom_phidot} are precisely the same as Eqs.~\Eqref{baseS2_thetadot} and \Eqref{baseS2_phidot} in Sec.~\ref{subsec_BaseS2}, and hence the solution may be fully described by Eq.~\Eqref{baseS2_soln}. As discussed in Sec.~\ref{subsec_BaseS2}, we may align the coordinate system such that $\theta$ is constant at $\cos\theta=-\frac{eB}{L}$ by choosing $K=L^2-e^2B^2$. Here we see that these quantities affect the $r$-motion through Eq.~\Eqref{fluxcomp_eom_rdot}, where we define the effective potential
\begin{align}
 \mathcal{U}=\brac{\frac{4K}{6B^2}+\frac{\Psi^2}{r^2}+1}f(r).
\end{align}
For instance, at fixed $L$, increasing the charge roughly lowers the graph of $\mathcal{U}$, as shown in Fig.~\ref{fig_fluxcomp_pot}.

Turning to the innermost stable circular orbits (ISCOs), we find them by the condition $\mathcal{U}'=\mathcal{U}''=0$. It turns out that the radius of the ISCO is the solution of 
\begin{align}
 24B^2m^2-4B^2mr+15mr^3-4r^4=0,
\end{align}
which is independent of $K$, and hence of $L$ and $e$. As $B$ determines the $\mathrm{AdS}_4$ curvature scale $\ell$ through $B^2=\frac{\ell^2}{2}$, the ISCO radius is the same as the ISCO of geodesics in $\mathrm{AdS}_4$ \cite{Berenstein:2020vlp}, and is unaffected by the particle's charge nor the motion in the $S^2$ directions.

\section{Magnetic fields with hyperbolic symmetry} \label{sec_hyperbolic}

Here we consider the case where $\mathcal{N}_2=H^2$, the hyperbolic manifold of constant negative curvature. In `pseudo-spherical' coordinates, its metric is
\begin{align}
 \bar{g}_{ab}\dif y^a\dif y^b=\dif\theta^2+\sinh^2\theta\,\dif\phi^2.
\end{align}
We take the electromagnetic two-form to be proportional to its volume form. That is,
\begin{align}
 F=B\sinh\theta\;\dif\theta\wedge\dif\phi.
\end{align}
This is the exterior derivative of the one-form potential $A=B\cosh\theta\;\dif\phi$.

The isometries of $H^2$ are
\begin{subequations} \label{C_hyp}
\begin{align}
 \xi_{(1)}&=-\sin\phi\partial_\theta-\coth\theta\cos\phi\partial_\phi,\\
 \xi_{(2)}&=\cos\phi\partial_\theta-\coth\theta\sin\phi\partial_\phi,\\
 \xi_{(3)}&=\partial_\phi.
\end{align}
\end{subequations}
Using Eq.~\Eqref{Qeqn2} to solve for $\Psi_{(i)}$, we obtain the constants of motion \cite{Comtet:1986ki}
\begin{subequations} \label{Q_hyp}
 \begin{align}
  Q_{(1)}&=-eB\sinh\theta\cos\phi-P_\theta\sin\phi-P_\phi\coth\theta\cos\phi,\\
  Q_{(2)}&=-eB\sinh\theta\sin\phi+P_\theta\cos\phi-P_\phi\coth\theta\sin\phi,\\
  Q_{(3)}&=eB\cosh\theta+P_\phi.
 \end{align}
\end{subequations}
Constants of motion for charged particle on hyperbolic plane have been found previously, such as in \cite{Comtet:1986ki} using other equivalent methods.

\subsection{Analysis of charged-particle motion on \texorpdfstring{$\mathcal{N}_2=H^2$}{N2=H2}} \label{subsec_BaseH2}

We first review the intrinsic problem of a (non-relativistic) charged particle moving on a hyperbolic plane $H^2$. This problem was considered in Refs.~\cite{Comtet:1984mm,Comtet:1986ki,2022arXiv220206055K}. Here we provide a detailed discussion using the constants of motion found in Eq.~\Eqref{Q_hyp}. Additionally, we will attempt to define the analogue of the Poincar\'{e} cone for this problem. 

In pseudo-spherical coordinates, the hyperbolic plane $H^2$ is described by the metric
\begin{align}
 \dif s^2&=a^2\brac{\dif\theta^2+\sinh^2\theta\,\dif\phi^2}.
\end{align}
The magnetic field along this plane is $F=B\sinh\theta\,\dif\theta\wedge\dif\phi$, where the magnetic potential is $A=B\cosh\theta\,\dif\phi$.

The Lagrangian for this problem is 
\begin{align}
 \mathcal{L}=\frac{a^2}{2}\brac{\dot{\theta}^2+\sinh^2\theta\,\dot{\phi}^2}+eB\cosh\theta\dot{\phi}.
\end{align}
The canonical momenta are
\begin{align}
 p_\theta=a^2\dot{\theta},\quad p_\phi=a^2\sinh^2\theta\dot{\phi}+eB\cosh\theta,
\end{align}
and the gauge-covariant momenta are 
\begin{align}
 P_\theta=p_\theta,\quad P_\phi=p_\phi-eB\cosh\theta.
\end{align}
Its corresponding Hamiltonian is $H=\frac{1}{2a^2}\brac{p_\theta^2+\frac{\brac{p_\phi-eB\cosh\theta}^2}{\sinh^2\theta}}$, from which we have the Hamilton--Jacobi equation
\begin{align}
 \frac{1}{2a^2}\sbrac{\brac{\frac{\partial S}{\partial\theta}}^2+\frac{1}{\sinh^2\theta}\brac{\frac{\partial S}{\partial\phi}-eB\cosh\theta}^2}+\frac{\partial S}{\partial\tau}=0.
\end{align}
The ansatz for $S$ is taken to be $S=-\half \mathcal{E}\tau+L\phi+S_\theta(\theta)$, where $\mathcal{E}$ is the (non-relativistic) total energy, $W=L\phi+S_\theta(\theta)$ as the Hamilton's characteristic function, and $L$ is the angular momentum associated with $\phi$-motion.

Letting $K=2a^2\mathcal{E}$, the equations of motion are 
\begin{align}
 \dot{\theta}&=\pm\sqrt{K-\frac{\brac{L-eB\cosh\theta}^2}{\sinh^2\theta}},\quad \dot{\phi}=\frac{L-eB\cosh\theta}{\sinh^2\theta}.
\end{align}
For solving the equations of motion, it is convenient to let $x=\cosh\theta$, where the equations of motion becomes
\begin{subequations}\label{BaseH2_eom}
 \begin{align}
  \dot{x}&=\pm\sqrt{X(x)},\label{hyp_xdot}\\
  \dot{\phi}&=\frac{L-eBx}{x^2-1},\label{hyp_phidot}\\
  X(x)&=(K-e^2B^2)x^2+2eBLx-K-L^2.\label{hyp_X}
 \end{align}
\end{subequations}
 The accessible domain of the particle are the points where $X(x)\geq0$. Here, $X(x)$ is a quadratic function where the coefficient of the quadratic term is $K-e^2B^2$. The roots of $X$ are
 \begin{align}
  x_\pm=\frac{eBL\pm\sqrt{K(K+L^2-e^2B^2)}}{e^2B^2-K}.
 \end{align}
We then have two cases. First, if $e^2B^2>K$, we have $x_-\leq x_+$ and $X(x)$ is non-negative at $x_-\leq x\leq x_+$. Hence the particle is bounded in this finite region. On the other hand, if $e^2B^2<K$, we have $x_-\geq x_+$. The function $X(x)$ is non-negative for $x\leq x_+$ and $x\geq x_-$. In this case the particle's motion is unbounded. We have essentially recovered the statements of Ref.~\cite{Comtet:1986ki}; the magnetic field has to be sufficiently strong to prevent the charged particle from escaping to infinity.
  
For a visual depiction of the particle motion, we wish, as in the spherical case, to embed the trajectory in an ambient three-dimensional space. However, by Hilbert's theorem a hyperbolic space cannot be embedded in a three-dimensional space of Euclidean signature. Instead, it is more natural to embed $H^2$ in a three-dimensional \emph{Minkowski} spacetime $\mathbb{R}^{2,1}$ with metric
\begin{align}
 \dif s^2=\brac{\dif X^1}^2+\brac{\dif X^2}^2-\brac{\dif X^3}^2=\eta_{ij}\dif X^i\dif X^j.
\end{align}
The embedding is achieved by the parametrisation
\begin{align}
 X^1&=a\sinh\theta\cos\phi,\quad X^2=a\sinh\theta\sin\phi,\quad X^3=a\cosh\theta.
\end{align}
Here $X^3$ is the time-like coordinate of this auxiliary Minkowski spacetime. 

The analogous notion of a `Poincar\'{e} cone' requires an appropriate modification to the $\mathbb{R}^{2,1}$ case. In the $S^2$ case, we defined it through the inner product in $\mathbb{R}^3$, where vectors are self-dual. Here, we should take the conserved quantities $Q_{(i)}$ as components of a \emph{dual vector}, or one-form 
\begin{align*}
 \mathcal{J}=Q_{(1)}\dif X^1+Q_{(2)}\dif X^2+Q_{(3)}\dif X^3.
\end{align*}
We define the `position vector' in $\mathbb{R}^{2,1}$ as
\begin{align}
 X=X^1\frac{\partial}{\partial X^1}+X^2\frac{\partial}{\partial X^2}+X^3\frac{\partial}{\partial X^3},
\end{align}
where the coordinates $X^i$ are used to defined the basis vectors $\frac{\partial}{\partial X^i}$, with the corresponding dual basis $\dif X^i$ such that
\begin{align}
 \left\langle\dif X^i,\frac{\partial}{\partial X^j}\right\rangle=\delta^i_j,
\end{align}
for $i,j=1,2,3$, and $\langle\cdot,\cdot\rangle$ denotes the inner product in $\mathbb{R}^{2,1}$. With this inner product we have 
\begin{align}
 \langle\mathcal{J},X\rangle=Q_{(i)}X^i=eBa, \label{hyp_InnerProd}
\end{align}
which is the hyperbolic counterpart to Eq.~\Eqref{sph_InnerProd}. 

This expression then gives the hyperbolic analogue of the Poincar\'{e} cone, the difference being $\mathbb{R}^{2,1}$ is carries a Lorentzian signature and we do not have a Euclidean geometric interpretation of a cone angle. 

\begin{figure}
 \centering 
 \includegraphics{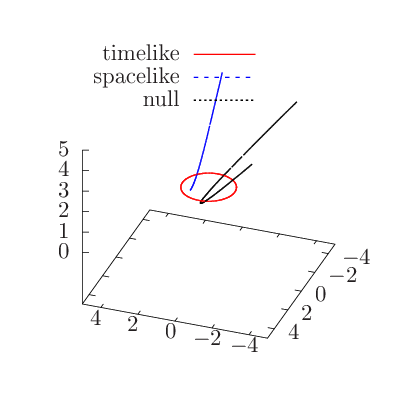}
 \caption{Motion of charged particles on $H^2$ with time-like, null, and space-like $J$. In the time-like case, $J$ is aligned to the `time' direction of $\mathbb{R}^{2,1}$, $X^3$, and an orbit is shown in red. For the null case $J$ is chosen such it is along the null surface $X^3=X^1$. In the space-like case, $J$ is aligned along the $X^1$ direction.}
 \label{fig_align}
\end{figure}

Our next question is whether it is possible to align our coordinate system to one of the coordinate axes, as we did in the $S^2$ case. As $SO(2,1)$ transformations preserves the inner product \Eqref{hyp_InnerProd}, there is a subtlety to be considered. Taking $Q^i=\eta^{ij}Q_{(j)}$ as components of the contravariant vector $J=Q^i\frac{\partial}{\partial X^i}$ whose dual is $\mathcal{J}$, we note that $Q^1$ and $Q^2$ are `spatial' components while $Q^3$ is `time-like' in the context of our auxiliary $\mathbb{R}^{2,1}$ Minkowski spacetime. Observe that 
\begin{align}
 Q^iQ_{(i)}=Q_{(1)}^2+Q_{(2)}^2-Q_{(3)}^2=-e^2B^2+K.
\end{align}
So the vector $Q=Q^i\frac{\partial}{\partial X^i}$ is time-like for bounded motion ($e^2B^2>K$), space-like for unbounded motion ($e^2B^2< K$), and null in the critical case ($e^2B^2=K$). By performing $SO(2,1)$ transformations, one may align $Q$ along a coordinate axis, just as we did in the spherical-symmetric case. However, such transformations preserves the time-like/null/space-like character of a vector. Hence, depending on the nature of $J$, we have three possibilities:
\begin{enumerate}
 \item If $J$ is time-like, it can be aligned along the $X^3$ `time' axis, so that 
 \begin{subequations}\label{Q_hyp_align}
\begin{align}
 Q_{(1)}=0&=-\cos\phi\brac{eB\sinh\theta+P_\phi\coth\theta}-P_\theta\sin\phi,\\
 Q_{(2)}=0&=-\sin\phi\brac{eB\sinh\theta+P_\phi\coth\theta}+P_\theta\cos\phi.
\end{align}
\end{subequations}
This can be achieved by the choice $K=e^2B^2-L^2$($<e^2B^2$), for which $x=\cosh\theta$ is constant at $x=\frac{eB}{L}$. For these values, we also have $\dot{\phi}=-L$ and Eq.~\Eqref{Q_hyp_align} is achieved. As a result, the motion is confined to a constant $x=\cosh\theta=\frac{eB}{L}$, which, in turn, correspond to $X^3$ being constant. An example of this motion is shown by the red circle in Fig.~\ref{fig_align}.

 \item If $J$ is null, we align it along the null direction $X^3=\pm X^1$, such that $Q_{(2)}=0$ and $Q_{(3)}=\pm Q_{(1)}$. This is achieved by fixing $K=e^2B^2$, and choosing initial conditions such that 
 \begin{subequations}
 \begin{align}
  P_\theta&=-\frac{eB\sin\phi}{\cosh\theta+\cos\phi\sinh\theta},\\
  P_\phi&=-\frac{eB\sinh\theta\brac{\sinh\theta+\cosh\theta\cos\phi}}{\cosh\theta+\cos\phi\sinh\theta}.
 \end{align}
 \end{subequations}
 An example of such a motion is shown by the black curve in Fig.~\ref{fig_align}.

 \item If $J$ is space-like, it can be aligned along the $X^1$, which is a spatial axis. This results in $Q_{(2)}=Q_{(3)}=0$. This is attained by choosing initial conditions such that 
 \begin{subequations}
 \begin{align}
  P_\theta&=-\frac{q\sin\phi}{\cos\phi\sinh\theta},\\
  P_\phi&=-eB\cosh\theta.
 \end{align}
 \end{subequations}
 This ensures $Q_{(2)}=Q_{(3)}=0$ throughout the motion. Therefore $\langle\mathcal{J}, X\rangle=Q_{(1)}X^1=eBa$ which means $X^1$ is constant throughout the motion. An example of such a motion is shown by the blue curve in Fig.~\ref{fig_align}.
\end{enumerate}
Therefore, in each of the time-like/null/space-like cases, there exist a submanifold in $\mathbb{R}^{2,1}$ which serves as the analogue of the Poincar\'{e} cone.
%

\begin{figure}
 \centering 
 \includegraphics{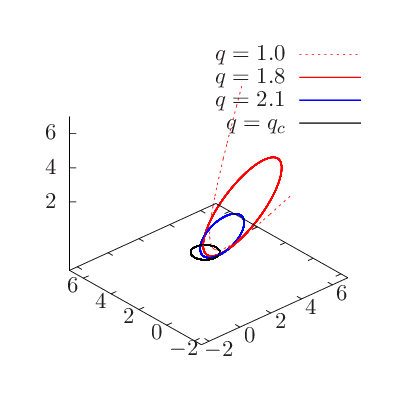}
 \caption{Motion of charged particle on $H^2$ with $K=2$, $L=2$ and various $q=eB$. Here $q_c=\sqrt{K+L^2}$.}
 \label{fig_Hcone}
\end{figure}

Turning now to the general analytic solution where $J$ is in an arbitrary direction, we choose $\phi=0$ and $x=x_-$ as the initial conditions and assuming $K+L^2-e^2B^2>0$, Eq.~\Eqref{BaseH2_eom} has the solution 
\begin{align}
 \phi(x)&=-\half\mathrm{sgn}(L-eB)\brac{\zeta(x)-\frac{\pi}{2}}+\half\mathrm{sgn}(L+eB)\brac{\eta(x)-\frac{\pi}{2}},\quad x_-<x\leq x_+,\\
 \zeta(x)&=\arcsin\left\{\frac{1}{\sqrt{K(K+L^2-e^2B^2)}}\sbrac{\frac{(L-eB)^2}{x-1}-\brac{K-e^2B^2+eBL}}\right\},\\
 \eta(x)&=\arcsin\left\{\frac{1}{\sqrt{K(K+L^2-e^2B^2)}}\sbrac{\frac{(L+eB)^2}{x+1}-(e^2B^2-K+eBL)}\right\}. \label{BaseH2_soln}
\end{align}
In the case of bounded motion, the sign functions `$\mathrm{sgn}$' determines whether the trajectory encloses $X^3$ axis. This is demonstrated in Fig.~\ref{fig_Hcone}, trajectories for $L=1$ and $a=\mathcal{E}=1$ are plotted for various $q=eB$. In this case, $K=2a^2\mathcal{E}=2$. The $X^3$-axis is vertical in this figure. The red dotted curve is the unbounded trajectory as $e^2B^2<K$.

\subsection{Is there a correspondence between charged particle motion with conical defect spacetimes?}

By comparison with the spherically symmetric case discussed in Sec.~\ref{subsec_correspondenceS2}, we wish to check whether there is an analogous correspondence between charged particle motion with geodesics in a conical defect spacetime. It turns out that the answer is in the negative. The main reason for this is the equations of motion for $\theta$ and $\phi$ are different when $\theta$ is non-constant. In the spherically-symmetric case, this can be acheived by aligning the Poincar\'{e} cone to the $X^3$-axis. However, geodesic motion in the hyperbolic case are unbounded in $\theta$, much less rendered constant. 

To see this more explicitly, suppose we wish to find a correspondence between the following two problems. The first is charged particle motion in the following solution containing hyperbolic symmetry:
\begin{subequations}
\begin{align}
 \dif s^2&=-\mathcal{A}\dif t^2+\mathcal{B}\dif r^2+\mathcal{C}\brac{\dif\theta^2+\sinh^2\theta\,\dif\phi^2},\\
 A&=B\cosh\theta\,\dif\phi,
\end{align}
\end{subequations}
where $\mathcal{A}$, $\mathcal{B}$, and $\mathcal{C}$ are functions of $r$. Separating the Hamilton--Jacobi equations for a charged particle results in the following equations of motion,
\begin{subequations}
 \begin{align}
  \dot{t}&=-\frac{E}{\mathcal{A}},\\
  \mathcal{B}\dot{r}^2&=\frac{E^2}{\mathcal{A}}-\frac{K}{\mathcal{C}}-1,\\
  \dot{\phi}&=\frac{L-eB\cosh\theta}{\mathcal{C}\sinh^2\theta},\label{corres1_phidot}\\
  \mathcal{C}^2\dot{\theta}^2&=K-\frac{\brac{L-eB\cosh\theta}^2}{\sinh^2\theta},\label{corres1_thetadot}
 \end{align}
\end{subequations}
where $E$ and $L$ are the conserved energy and angular momentum, respectively, and $K$ is the Carter-like separation constant. 

Following the lines of Eq.~\Eqref{subsec_correspondenceS2}, we seek a correspondence to a second problem which is geodesic motion in the spacetime
\begin{align}
 \dif s^2&=-\mathcal{A}\dif t^2+\mathcal{B}\dif r^2+\mathcal{C}\brac{\dif\theta^2+\eta^2\sinh^2\theta\,\dif\phi^2},
\end{align}
where a conical defect is present for $\eta\neq 1$. However, the separated Hamilton--Jacobi equation for this case is 
\begin{subequations}
 \begin{align}
  \dot{t}&=-\frac{\mathcal{E}}{\mathcal{A}},\\
  \mathcal{B}\dot{r}^2&=\frac{\mathcal{E}^2}{\mathcal{A}}-\frac{\mathcal{K}}{\mathcal{C}}-1,\\
  \dot{\phi}&=\frac{\Phi}{C\eta^2\sinh^2\theta},\label{corres2_phidot}\\
  \mathcal{C}^2\dot{\theta}^2&=\mathcal{K}-\frac{\Phi^2}{\eta^2\sinh^2\theta},\label{corres2_thetadot}
 \end{align}
 where $\mathcal{E}$ and $\Phi$ are the conserved energy and angular momentum, respectively, and $\mathcal{K}$ is the Carter-like separation constant.
 
 We see that Eq.~\Eqref{corres1_phidot} can be made equivalent to \Eqref{corres2_phidot} by an appropriate identification of conserved quantities only if $\theta$ is constant. However, letting $x=\cosh\theta$, we see that $\dot{\theta}=0$, or equivalently, $\dot{x}=0$ if
 \begin{align}
  x_\pm=\pm\sqrt{1+\frac{\Phi^2}{K\eta^2}}.
 \end{align}
 When $x_+$ and $x_-$ are distinct, the point $\dot{\theta}=0$ is only an instantaneous turning point, and $\theta$ does not remain constant throughout the motion. The two roots are degenerate when both are equal zero, $x_\pm=0$. However, since $x=\cosh\theta$, this cannot hold for real $\theta$. 
 
 Physically, this reflects the fact that geodesic motion on a hyperbolic spactime is unbounded. For charged particles in a magnetic field, we have seen above (and in \cite{Comtet:1986ki}) that a sufficiently strong $eB$ is required to prevent the particle from escaping to infinity. Below this threshold, $\theta$ cannot be made constant. From these, we conclude that there is no correspondence between charged-particle motion and geodesics in the conical defect spacetime, at least in the form analogous to the spherical case.

\end{subequations}

\subsection{Charged particle in Minkowski spacetime with a hyperbolic magnetic field}

As a simple model with a hyperbolic magnetic field, consider the four-dimensional Minkowski spacetime $\mathbb{R}^{1,3}$ with the metric in standard Minkowski coordinates:
\begin{align}
 \dif s^2&=-\dif t^2+\dif x^2+\dif y^2+\dif z^2.
\end{align}
We introduce a hyperbolic foliation of this spacetime by writing 
\begin{align}
 t=\sigma\cosh\theta,\quad x=\sigma\sinh\theta\cos\phi,\quad y=\sigma\sinh\theta\sin\phi,\quad z=z.
\end{align}
(Note that this parametrisation does not cover the whole Minkowski spacetime.) The  metric now becomes
\begin{align}
 \dif s^2&=-\dif\sigma^2+\sigma^2\brac{\dif\theta^2+\sinh^2\theta\,\dif\phi^2}+\dif z^2. \label{boost_Mink}
\end{align}
For this problem, we assume the magnetic field is a test field that does not backreact to the curvature of spacetime. The magnetic field arising from the potential $A=B\cosh\theta\,\dif\phi$ solves the Maxwell's equation with \Eqref{boost_Mink} as the background.
 
To obtain the equations of motion for the particle, we turn to its Hamilton--Jacobi equation
\begin{align}
 \half\sbrac{-\brac{\frac{\partial S}{\partial\sigma}}^2+\brac{\frac{\partial S}{\partial z}}^2+\frac{1}{r^2}\brac{\frac{\partial S}{\partial\theta}}^2+\frac{1}{\sigma^2\sinh^2\theta}\brac{\frac{\partial S}{\partial\phi}-eB\sinh\theta}^2}+\frac{\partial S}{\partial\tau}=0.
\end{align}
Taking the ansatz $S=\half\tau+\eta z+L\phi+S_\sigma(\sigma)+S_\theta(\theta)$, then we find a separation constant $K$, leading to 
\begin{subequations}
\begin{align}
 \dot{\sigma}&=\pm\sqrt{\eta^2+1+\frac{K}{\sigma^2}},\label{boost_Mink_sigmadot}\\
 \sigma^2\dot{\theta}&=\pm\sqrt{K-\frac{(L-eB\cosh\theta)^2}{\sinh^2\theta}},\label{boost_Mink_thetadot}\\
 \sigma^2\dot{\phi}&=\frac{L-eB\cosh\theta}{\sinh^2\theta},\label{boost_Mink_phidot}\\
 \dot{z}&=\eta.\label{boost_Mink_zdot}
\end{align}
\end{subequations}

Eliminating $\sigma$ between \Eqref{boost_Mink_thetadot} and \Eqref{boost_Mink_phidot} leads to an equation identical to \Eqref{BaseH2_eom}. Hence the solution \Eqref{BaseH2_soln} can be carried over. The other coordinates $\sigma$ and $z$ have simple solutions which increase linearly with $\tau$.

\subsection{Magnetic hyperbolic Reissner--Nordstr\"{o}m-AdS spacetime}

We now briefly consider charged particle around a magnetic Reissner--Nordstr\"{o}m black hole with a hyperbolic horizon \cite{Banados:1997df,Vanzo:1997gw,Mann:1997iz,Birmingham:1998nr,Mann:1997jb,Birmingham:1998nr}, described by 
\begin{align}
 \dif s^2&=-f(r)\dif t^2+f(r)^{-1}\dif r^2+r^2\brac{\dif\theta^2+\sinh^2\theta\,\dif\phi^2},\\
 A&=B\cosh\theta\,\dif\phi,\quad f(r)=-1-\frac{2m}{r}+\frac{B^2}{r^2}+\frac{r^2}{\ell^2}.
\end{align}
This metric and magnetic field solves the Einstein--Maxwell equations with a negative cosmological constant $\Lambda=-3/\ell^2$. The solution for $\Lambda=0$ was also considered by \cite{Herrera:2018mzq,Herrera:2020bfy} as a static model depicting the interior of a black hole.

For a charged particle in this spacetime, the Hamilton--Jacobi equation is 
\begin{align}
 \half\sbrac{-\frac{1}{f}\brac{\frac{\partial S}{\partial t}}^2+f\brac{\frac{\partial S}{\partial r}}^2+\frac{1}{r^2}\brac{\frac{\partial S}{\partial\theta}}^2+\frac{1}{r^2\sin^2\theta}\brac{\frac{\partial S}{\partial\phi}-eB\cosh\theta}^2}+\frac{\partial S}{\partial\tau}&=0.
\end{align}
With the usual ansatz, we find the separated equations of motion 
\begin{subequations}
 \begin{align}
  \dot{t}&=-\frac{E}{f},\label{RNAdS_tdot}\\
  \dot{r}&=\pm\sqrt{E^2-\brac{\frac{K}{r^2}+1}f},\label{RNAdS_rdot}\\
  r^2\dot{\theta}&=\pm\sqrt{K-\frac{(L-eB\cosh\theta)^2}{\sinh^2\theta}},\label{RNAdS_thetadot}\\
  r^2\dot{\phi}&=\frac{L-eB\cosh\theta}{\sinh^2\theta}.\label{RNAdS_phidot}
 \end{align}
\end{subequations}
Eliminating $r$ between \Eqref{RNAdS_thetadot} and \Eqref{RNAdS_phidot} we obtain an equation identical to \Eqref{BaseH2_eom}, so the exact solution in \Eqref{BaseH2_soln} can also be carried over.

\begin{figure}
 \centering
 \begin{subfigure}[b]{0.49\textwidth}
    \centering
    \includegraphics[width=\textwidth]{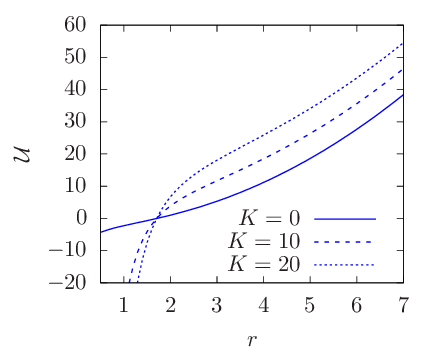}
    \caption{$m=1$.}
    \label{fig_HyperRN_Ua}
  \end{subfigure}
  \begin{subfigure}[b]{0.49\textwidth}
    \centering
    \includegraphics[width=\textwidth]{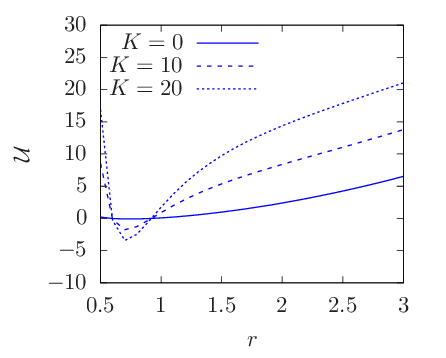}
    \caption{$m=0$.}
    \label{fig_HyperRN_Ub}
  \end{subfigure}
  \begin{subfigure}[b]{0.49\textwidth}
    \centering
    \includegraphics[width=\textwidth]{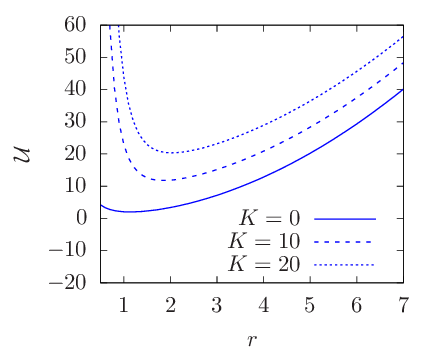}
    \caption{$m=-1$.}
    \label{fig_HyperRN_Uc}
  \end{subfigure}
  \caption{$\ell^2=1.2$, $B=0.5$, $L=1$, and various $K$.}
  \label{fig_HyperRN_U}
\end{figure}

This is determined by Eq.~\Eqref{RNAdS_rdot}, from which we define an effective potential 
\begin{align}
 \mathcal{U}=\brac{\frac{K}{r^2}+1}f,
\end{align}
Fig.~\ref{fig_HyperRN_U} shows various the effective potential for various $K$ and $m$. We see that bounded motion in the $r$-direction are typically possible for black holes of negative mass (For instance, Fig.~\ref{fig_HyperRN_Uc} for $m=-1$.)

\begin{figure}
 \centering 
 \includegraphics{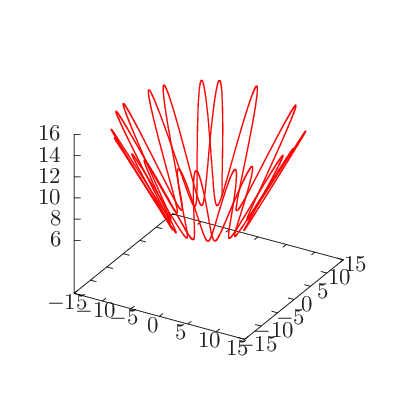}
 \caption{Charged particle in the magnetic, hyperbolic RN-AdS spacetime with $m=-1$, $B=0.5$, $\ell^2=1$, $L=1$, $K=20$, $eB=\sqrt{K+L^2}$, and $E=\sqrt{30}$.}
 \label{fig_HyperRN_orbit}
\end{figure}

To visualise the trajectory, we again make use of the embedding in $\mathbb{R}^{2,1}$ with
\begin{align}
 X^1=r\sinh\theta\cos\phi,\quad X^2=r\sinh\theta\sin\phi,\quad X^3=r\cosh\theta.
\end{align}
So the trajectory is shown as a curve in auxiliary Minkowski spacetime with $X^3$ as its time direction. For instance, Fig.~\ref{fig_HyperRN_orbit} shows an example of a bounded orbit with $K=e^2B^2-L^2$. By the discussion in Sec.~\ref{subsec_BaseH2}, this choice of $K$ aligns the hyperbolic `Poincar\'{e} cone' along the $X^3$ axis. In the present case, there is motion along $r$ as well. Hence the trajectory lies on the surface of a cone of constant $\theta$.

\section{Conclusion} \label{sec_conclusion}

In this paper we have considered charged particle motion in various spacetime with magnetic fields of spherical and hyperbolic symmetry. Using Poisson-bracket methods, constants of motion associated to the spherical and hyperbolic symmetry are found using van Holten's procedure \cite{vanHolten:2006xq}. The equations of motion along the $S^2$ and $H^2$ directions are solved exactly for the general case. 

In the spherical case, trajectories are visualised in an auxiliary flat space $\mathbb{R}^3$. A conserved vector $\vec{J}$ can be constructed out of the constants of motion. The Poincar\'{e} cone is then defined through the dot product with this vector. We have also established a correspondence between charged-particle motion on the Poincar\'{e} cone with geodesics on related spacetimes with a conical defect. We demonstrate this correspondence with a Schwarzschild spacetime with a test magnetic monopole field, whose charged particle motion is in correspondence with geodesic motion around a Schwarzschild black hole threaded by a cosmic string. 

In the hyperbolic case, we also have three constants of motion. Instead of Euclidean flat space, trajectories are more naturally visualised with an auxiliary three-dimensional Minkowski spacetime $\mathbb{R}^{2,1}$. The constants of motion can then be used to define a conserved vector in $\mathbb{R}^{2,1}$, where it is time-like when the motion is bounded and space-like or null when the motion is unbouned. With this vector we have attempted to define the analogue of the Poincar\'{e} through the inner product under the Minkowski metric of $\mathbb{R}^{2,1}$.

\section*{Acknowledgments}
 
Y.-K.~L is supported by Xiamen University Malaysia Research Fund (Grant no. XMUMRF/ 2021-C8/IPHY/0001).

\bibliographystyle{cones}

\bibliography{cones}

\end{document}